# Weighted H-index for identifying influential spreaders

Senbin Yu[1], Liang Gao[1, 2, *], Yi-Fan Wang[1], Ge Gao[1], Congcong Zhou[1], and Zi-You Gao[1]

1. Institute of Transportation Systems Science and Engineering, MOE Key Laboratory of Urban Transportation System Theory and Technology, State Key Laboratory of Rail Traffic Control and Safety, and Center of Cooperative Innovation for Beijing Metropolitan Transportation, Beijing Jiaotong University, Beijing 100044, China
2. Yinchuan Municipal Bureau of Big Data Management and Service, Yinchuan 750011, P. R. China

**Abstract**：Spreading is a ubiquitous process in the social, biological and technological systems. Therefore, identifying influential spreaders, which is important to prevent epidemic spreading and to establish effective vaccination strategies, is full of theoretical and practical significance. In this paper, a weighted h-index centrality based on virtual nodes extension is proposed to quantify the spreading influence of nodes in complex networks. Simulation results on real-world networks reveal that the proposed method provides more accurate and more consistent ranking than the five classical methods. Moreover, we observe that the monotonicity and the computational complexity of our measure can also yield excellent performance.

## 1. Introduction

Many spreading phenomena like the cascading failures [1], virus transmission [2], rumors diffusion [3] and so forth in the real word can be described as the spreading process on the complex networks. The understanding of significant role that a single node plays provides pregnant insights into network structure and functions [4]. So identifying influential spreaders in complex networks has increases much attention. And the fundamental problem is how to identifying and ranking the efficient spreaders in this research.

Degree centrality, the simplest indicator, focuses on number of links per node believes that the most connected nodes are hubs [5]. There are also many classic topology metrics such as betweenness centrality [6], closeness centrality [7] and Katz centrality [8]. These measures show good performance in distinguishing different influential nodes, but the computational complexity is unacceptable when they are applied to large-scale networks. While Kitsak et al. [9] argued that the most influential spreaders are the nodes reside in the core of the network by the k-shell decomposition analysis. However, the k-shell decomposition tends to assign many nodes with different spreading ability in the same coreness index [10]. Thus, researchers proposed some improved methods to overcome this shortcoming. For example, Zeng et al. proposed a mixed degree decomposition (MDD) method by considering both the residual degree and the exhausted degree [11], but the optimal parameter $\lambda$ is uncertain. Liu et al. took into account the shortest distance from a target node to the node of highest k-shell values and presented a more distinguishable ranking list [1_0]. To evaluate the spreading influence of a node, Bae and Kim proposed a novel measure called neighborhood coreness centrality $C_{nc+}$ which used the neighborhood coreness of neighbor [16]. Wang et al. utilized the iteration information produced in k-shell decomposition and presented a ranking method to evaluate the influence capability of nodes [13]. Shuang et al. designed an iterative neighbor information gathering (ING) process to rank the node influence [18]. Besides, there are some other node ranking algorithms have been introduced to achieve the promotion of ranking performance [14, 15, 17, 1_1, 1_2, 1_3, 1_7, 1_6]. Except for the improved k-shell decomposition algorithm, there existed many other excellent algorithms. Liu et al. [1_4] used the total asymmetric link weights to quantify the impact of the node in spreading processed. Zhang et al. [1_5] proposed a VoteRank method to identify a set of decentralized spreaders with the best spreading ability. Ma et al. [1_8] modified local centrality and integrate it with DC by considering the spreading probability.

In this paper, we argue that edges in a network could be quite different [1_9] and have different significance in network structure and function [1_10]. Many measure use the edge's importance to define the importance of a node [1_4, 1_12, 1_13, 1_14, 1_15]. For example, the number of shortest paths go through the edge and it can be regarded as the weight of an edge [1_11]. A measure $d_{ij} = (d_i * d_j)^a$ [1_16], which was found to correlate positive with the volume of passengers traveling between two airports, has been adopted in many works for making a distinction among edges in unweighted networks. Recently Lü et al. [1_17] constructed an operator $\mathcal{H}$ on the neighbor's degree of a node and obtained an h-index of each node. The h-index was the overall best in performers when compared with three typical centralities for undirected networks. Inspired by these factors, we propose a weight edge by the product of two degrees of connected nodes. And then utilize the operator $\mathcal{H}$ on the neighbors of each node which are extended by $k$ weight edges, where $k$ is the degree of each neighbor. The sum of neighbors' weighted h-index values defines the importance of a node. To evaluate the effectiveness of the proposed measure, we apply the susceptible-infected-recovered (SIR) model for investigating an epidemic spreading process on sixteen real-world networks. The results show that the proposed method has a better performance of ranking the spreading ability of nodes in general than five other centralities, which compared by making a rank correlation between ranking lists of centrality measure and simulation results by SIR model. Moreover, calculating the weight h-index centrality has a complexity of $O(m)$, where $m$ is the number of edges. The weighted h-index centrality is more efficient than other time consuming measure such as betweenness centrality and closeness centrality.

The remainder of this paper is organized as followers. We review the definition of centrality measures used for comparison and introduce our method in section 2. Section 3 reports evaluation methodologies and the experimental results. The conclusion are presented in section 4.

## 2. Centrality measures

In this part, we introduce the classic centrality measures and the proposed method. For a given unweighted complex network $G = (V, E)$, $n = |V|$ is the number of nodes, and $m = |E|$ is the number of edges. Let $e_{ij}$ be the value of edge if node $i$ is connected to node $j$. And we use $\varGamma_i$ to denote the set of neighbors of node $i$.

## 2.1 Degree centrality

The degree centrality (D) is the simplest indicator to quantify node importance. It focuses on number of links per node and believes that the most connected nodes are hubs. Let $D(i)$ denote the D of node $i$, which is defined as:

$$D(i) = \frac{d_i}{m}, \tag{1}$$

where $d_i$ is the degree of node $i$.

## 2.2 Betweenness centrality

The betweenness centrality (B) of a node $i$ is the sum of the fraction of all-pairs that pass through node $i$. We set $B(i)$ as the B of vertex $i$ which is given by:

$$B(i) = \frac{s(i)}{S}, \tag{2}$$

where $s(i)$ and $S$ represent the number of shortest paths pass node $i$ and the sum of shortest paths in a graph, respectively. A node with higher B will have more control over the network, because more information will pass through this node like in a telecommunications network.

## 2.3 Closeness centrality

The closeness centrality (C) of a node is a measure of centrality in a network, calculated as the sum of the length of the shortest paths between the node and all other nodes in the graph. Thus the more central a node is, the closer it is to all other nodes. The C of node $i$ is defined as:

$$C(i) = \frac{1}{\sum_{j \in V} s_{ij}}, \tag{3}$$

where $s_{ij}$ is the shortest distance between node $i$ and node $j$.

## 2.4 K-shell centrality

The k-shell centrality (KS) is obtained in the k-shell decomposition process. Each node will be assigned to a k-shell index by the process recursively pruning nodes with degree less than or equal to $k$. The pruning process continues until all nodes in the network are removed. As a result, each node is associated with one k-shell index.

## 2.5 H-index

The h-index (H) of node $i$ in a network is defined as the maximum value $h$ such that there are at least $h$ neighbors of degree larger than or equal to $h$. It is an operator acts on a finite number of integer $(k_1, k_2, \cdots, k_j)$ and return an h-index value of node $i$, where $k_1, k_2, \cdots, k_j$ is the degree of neighbor nodes. Hence, we set $h_i$ as the h-index of node $i$ as follow:

$$h_i = \mathcal{H}(\{k_j\}_{j \in \Gamma_i}), \tag{4}$$

where $k_j$ is the degree of neighbor node $i$.

**2.6 Weighted H-index**

The edges in an unweighted network are treated as a same value. In fact, the edges have different significance in network structure and function. We define the edge weights by degree to quantify the diffusion capacity of links.

**Definition I.** The weight of edge $w_{ij}$ is defined as:

$$w_{ij} = k_i * k_j, \tag{5}$$

where $k_i$ and $k_j$ are the degree of node $i$ and node $j$ respectively, if node $i$ connect with node $j$ directly. The edges' weights counted by expression (5) in a diagram of a network is shown in Fig. 1.

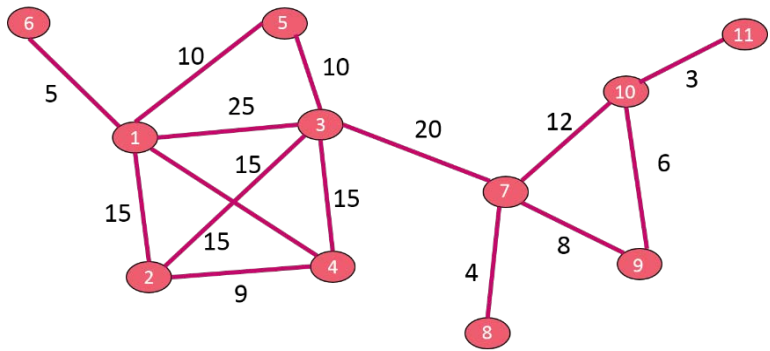

**Fig. 1.** A schematic representation of a network.

**Definition II.** In order to apply $\mathcal{H}$ operation on a weighted network, we decompose a weighted edge into multiple weighted edges. We take Fig. 2 as an example to illustrate the computational procedure.

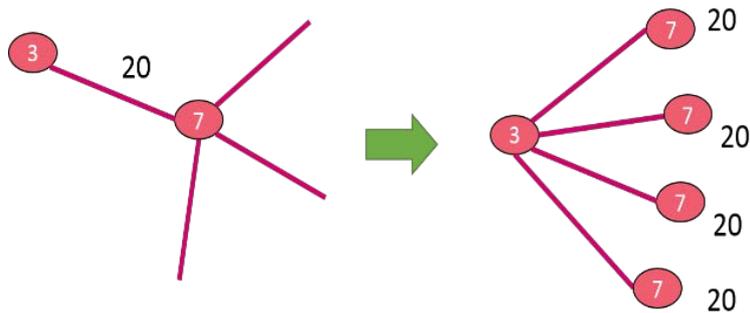

**Fig. 2.** The decomposition process of the weighted edge between node 3 and node 7.

We focus on node 3 if we count the importance of node 3. The edge $e_{37}$ is one of edges connected to node 3 and the weight of $e_{37}$ is 20 measured by formula (5). The $w_{37}$ can be extended $w_{37}$ into 4 weighted edges based on the degree of node 7. Then the process will continue until all node 3's neighbor nodes being counted. The expression (4) can write as: $\mathcal{H}\{\{25,\cdots,25\},\{20,\cdots,20\},\{15,15,15\},\{15,15,15\},\{10,10\}\}$. Final, we get the weighted h-index of node 3: $wh_3 = 15$. So

the formula (4) can be modified as follow:

$$wh_i = \mathcal{H}(\{\{w_{ij,1}, w_{ij,2}, \cdots, w_{ij,n}\}\}_{j \in \Gamma_i, n=k_j}), \tag{6}$$

where $w_{ij,1} = w_{ij,2} = \cdots = w_{ij,n} = w_{ij}$ and $k_j$ is degree of node $j$.

The $h_1$ and $h_3$ are 3. And we can get $wh_3 = 15$ and $wh_1 = 12$. It is worth mentioning that the spreading ability of node 1 and node 3 are 0.30963 and 0.30145 simulated in the SIR model, respectively. So the weighted h-index can distinguish the diffusion importance of node and rank them correctly.

**Definition III.** We define the h-index centrality (WH) of node $i$ as follow:

$$WH(i) = \sum_{v_j \in \Gamma(i)} wh_j. \tag{7}$$

In Table 1, we show the values measured by six methods for each nodes in sketch map. Using the spreading probability $\beta = 0.29 > \beta_c = 0.28846$, where $\beta_c$ is the epidemic threshold of the network, the spreading ability $S_\beta$ of each node is simulated in the SIR model. When we observe the ranking of node 7 and 5 by weighed h-index, they are in the reverse order to the spreading ability $S_\beta$. But this situation has been corrected by the measure of WH. The spreading ability of nodes identified by WH is unanimous completely with the simulation results. But the other centralities have more or less inconsistent ordering in their rankings. So we can see that the proposed measure can better rank the spreading ability of nodes than other centrality method considered.

Table 1 The centrality values of degree centrality (D), betweenness centrality (B), closeness centrality (C), k-shell centrality (KS), h-index (H) and weighted h-index centrality (WH) is estimated at $\beta = 0.29 > \beta_c = 0.28846$ by SIR model per seed node, where $\beta_c \sim \langle k \rangle / \langle k^2 \rangle$ is the epidemic threshold.

| Node | $S_\beta$ | D | B | C | KS | H | WH |
| --- | --- | --- | --- | --- | --- | --- | --- |
| 3 | 0.30963 | 0.5 | 0.57778 | 0.625 | 3 | 3 | 53 |
| 1 | 0.30145 | 0.5 | 0.22222 | 0.5 | 3 | 3 | 52 |
| 4 | 0.26509 | 0.30000 | 0.0 | 0.45455 | 3 | 3 | 38 |
| 2 | 0.26326 | 0.30000 | 0.0 | 0.45455 | 3 | 3 | 38 |
| 7 | 0.26000 | 0.4 | 0.6 | 0.58824 | 2 | 3 | 31 |
| 5 | 0.23818 | 0.2 | 0.0 | 0.43478 | 2 | 2 | 27 |
| 9 | 0.19327 | 0.2 | 0.0 | 0.41667 | 2 | 2 | 15 |
| 10 | 0.21854 | 0.30000 | 0.2 | 0.43478 | 2 | 2 | 18 |
| 6 | 0.16654 | 0.1 | 0.0 | 0.34483 | 1 | 1 | 12 |
| 8 | 0.15436 | 0.1 | 0.0 | 0.38462 | 1 | 1 | 9 |
| 11 | 0.14527 | 0.1 | 0.0 | 0.3125 | 1 | 1 | 6 |

## 3. Experimental results

In this section, we evaluate the effectiveness and the efficiency of proposed measure. Sixteen real-networks: four social networks (Karate club [25], Facebook [26], PGP [27] and Dolphins [28]), two communication networks (Enron [29] and Email [30]), six collaboration networks (AstroPh [31], CondMat [31], GrQc [31], HepPh [31], HepTh [31] and Jazz [32]), an autonomous system peering information from Oregon route-views (Oregon [33]), a transportation network (USAir [35]), a

C.elegans metabolic network (C.elegans [36]) and a power grid network (Power Grid [37]) listed in Table 2 are examined. And their basic topological features are summarized in the table.

Table 2. Properties of the sixteen real-world networks studies in this work. Structural properties of different networks include number of nodes $(n)$, number of edges $(m)$, epidemic threshold ($\beta_c \sim \langle k \rangle / \langle k^2 \rangle$), infection probability $(\beta)$ in the SIR spreading in the main text, average degree ($\langle k \rangle$), maximum degree ($k_{max}$), degree assortativity $(r)$, clustering coefficient $(C)$ and maximum k-shell index ($KS_{max}$).

| Network | $n$ | $m$ | $\beta_c$ | $\beta$ | $\langle k \rangle$ | $k_{max}$ | $r$ | $C$ | $KS_{max}$ |
|---|---|---|---|---|---|---|---|---|---|
| Karate | 34 | 78 | 0.129 | 0.13 | 4.5882 | 17 | -0.4756 | 0.5706 | 4 |
| Dolphins | 62 | 159 | 0.147 | 0.148 | 5.1290 | 23 | -0.04359 | 0.2590 | 4 |
| Jazz | 198 | 2742 | 0.0258 | 0.026 | 27.6970 | 100 | 0.02023 | 0.6175 | 29 |
| USAir | 332 | 2126 | 0.0225 | 0.023 | 12.8072 | 139 | -0.2079 | 0.6252 | 26 |
| C. | 453 | 2025 | 0.0249 | 0.025 | 8.9404 | 237 | -0.2258 | 0.6465 | 10 |
| E-mail | 1133 | 5451 | 0.0535 | 0.054 | 9.6222 | 71 | 0.07820 | 0.2202 | 11 |
| Faceboo | 4039 | 88234 | 0.0093 | 0.0094 | 43.6910 | 1045 | 0.06358 | 0.6055 | 115 |
| Power | 4941 | 6594 | 0.2583 | 0.26 | 2.66910 | 19 | 0.00345 | 0.0801 | 5 |
| GrQc | 5242 | 14496 | 0.0593 | 0.06 | 5.5307 | 81 | 0.6593 | 0.5296 | 43 |
| HepTh | 9877 | 25998 | 0.0798 | 0.08 | 5.2644 | 65 | 0.2678 | 0.4714 | 31 |
| PGP | 1068 | 24316 | 0.0529 | 0.053 | 4.5536 | 205 | 0.2382 | 0.2659 | 31 |
| Oregon | 1072 | 21999 | 0.0039 | 0.004 | 4.1009 | 2315 | -0.1889 | 0.2921 | 15 |
| HepPh | 1200 | 11852 | 0.0076 | 0.0077 | 19.7403 | 491 | 0.6323 | 0.6115 | 238 |
| AstroPh | 1877 | 19811 | 0.0153 | 0.016 | 21.1070 | 504 | 0.2051 | 0.6306 | 56 |
| CondMat | 2313 | 93497 | 0.0453 | 0.046 | 8.083 | 279 | 0.1340 | 0.6334 | 25 |
| Enron | 3669 | 18383 | 0.0071 | 0.007 | 10.020 | 1383 | -0.1108 | 0.4970 | 53 |

**3.1 Evaluation methodologies**

To study the spreading process, we use the SIR model to investigate the correctness of difference measures. In the initial time, there is only one infected seed node (I) and all other nodes are susceptible state (S). At each time step, infected nodes attempt to infect their susceptible neighbors with a probability $\beta$ and then enter the recovered state (R). This process is repeated until there are no longer any infected nodes. In our simulation, we use relatively small values, so the infected percentage of population is small. When the values of $\beta$ are high, any originated node can infect a large percentage of the population, the important of an individual node cannot be measured. Based on the heterogeneous mean-field theory [21, 22, 23], we set the infection probability $\beta$ to be slightly greater than the epidemic threshold $\beta_c \sim \langle k \rangle / \langle k^2 \rangle$.

In order to evaluate the correctness of the different methods, we adopt Kendall's tau coefficient as a rank correlation coefficient. In statistics, the Kendall's tau coefficient is used to measure the ordinal association between two measured quantities [24]. The Kendall's tau coefficient of two rank vectors $R$ (ranking method $\theta$) and $\sigma$ (SIR model) defined as:

$$\tau(R_\theta, \sigma) = \frac{n_c - n_d}{\sqrt{(n_0 - n_1)(n_0 - n_2)}}, \tag{8}$$

where $n_c$ and $n_d$ are the number of concordant pairs and discordant pairs respectively and $n_0 = n(n-1)/2$, $n_1 = \sum_i t_i (t_i - 1)/2$, $n_2 = \sum_j u_j (u_j - 1)/2$, where $n$ is the size of rank vectors and $t_i$ and $u_j$ are the number of tied values in the $i_{th}$ and $j_{th}$ group of ties, respectively.

We use the imprecision function, which is initially proposed by Kitsak et al. [9], to quantify the accuracy in pinpointing the most influential spreaders. This function is used to measure the difference between the average spreading of the $pN$ nodes with highest values by $\theta$ and the spreading of the $pN$ most efficient spreaders according to SIR dynamics. The imprecision function is

$$\varepsilon_\theta(p) = 1 - \frac{M_\theta(p)}{M_{\text{eff}}(p)}. \quad (9)$$

where $p$ is the fraction of network size $N$, $M_\theta(p)$ and $M_{\text{eff}}(p)$ are the average spreading efficiency of $pN$ nodes carrying the highest values and the highest actual spreading efficiency according to the simulated results of SIR model. A smaller $\varepsilon_\theta$ represents a higher accuracy of $\theta$ in identifying the most influential spreaders.

The monotonicity $M$ of ranking vector $R$ described in Ref [16] is adopted to quantify the resolution of different ranking measures as follow:

$$M(R) = (1 - \frac{\sum_{r \in R}|u_r|(|u_r|-1)}{|U|(|U|-1)})^2, \quad (10)$$

where $|U|$ is the ranking number of vector $R$, $R$ denotes the ranking vector of network nodes, and $|u_r|$ represents the number of ties with the same rank $r$. The monotonicity $M(R)$ is 1 if vector $R$ is perfectly monotonic, and it becomes 0 if all nodes are same in vector $R$.

## 3.2 Correctness

### 3.2.1 The performance of impression function

Figure 3 shows that the impression function of ranking based on the proposed method and the other five centralities. Recall that a lower impression implies a high accuracy in identifying the most influential spreaders. We can see that WH (gold circles) give an imprecision that is less than 0.1 for all $p$ ranging from 0.01 to 0.30 in nearly all cases. Only in the network Karate Club and HePth do the imprecisions get close to 0.20 when $p$ is near 0.02. More noticeable is that the WH performs even better than the other five measures in most case, except at some smaller values of $p$ in few parts of the sixteen networks. The B (blue stars) and C (red xs) are always the worst performance in the measure of impression function. The H (green triangles) is regarded as a trade-off between degree and coreness and has a best performers among D, KS and H centrality, although they are pretty close in most subfigures of Figure 3. The impression function demonstrate the improved performance of WH in identifying the most influential spreaders.

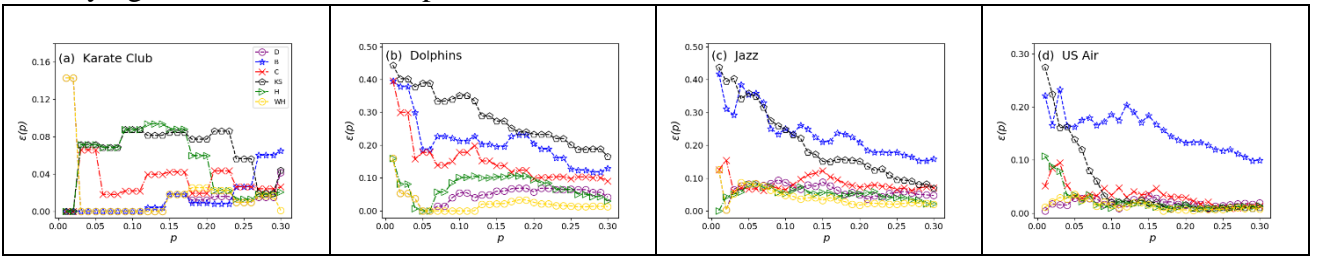

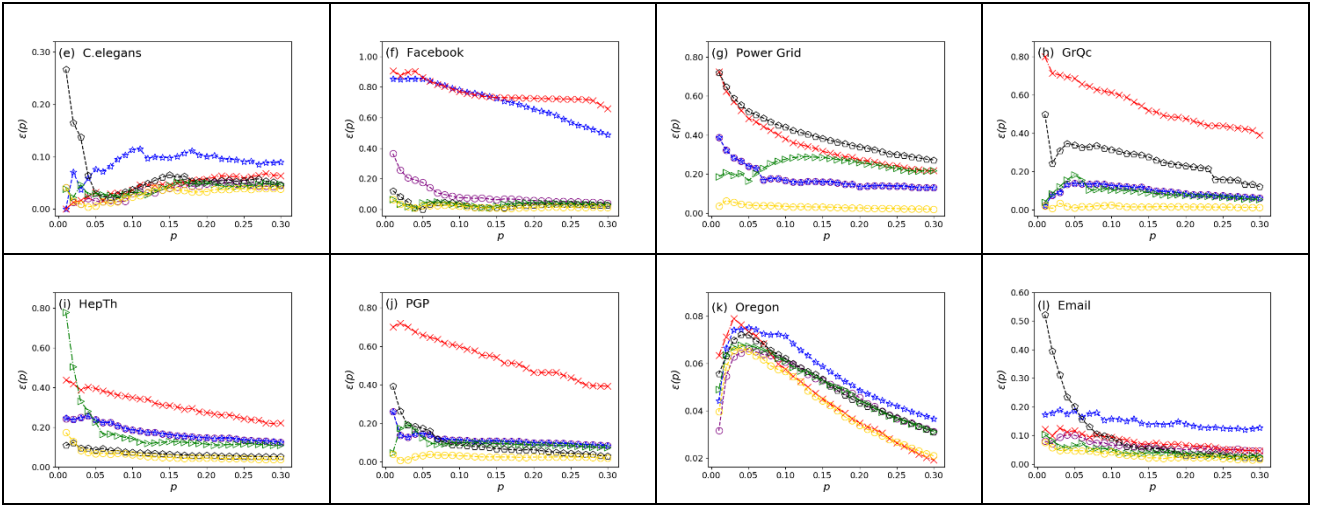

Fig. 3. The imprecision functions $\varepsilon_\theta(p)$ test using degree centrality (D), betweenness centrality (B), closeness centrality (C), k-shell centrality (KS), H-index (H) and the proposed centrality (WH) in the twelve real-world networks. By ranging from 0.01 to 0.30, $p$ is the proportion of nodes. The real-world networks are (a) Karate Club, (b) Dolphins, (c) Jazz, (d) US Air, (e) C.elegans, (f) Facebook, (g) Power grid, (h) GrQc, (i) HepTh, (j) PGP, (k) Oregon, (l) Email

The WH method yields consistently lower imprecision compared to the benchmark methods.

### 3.2.2 Comparison of rank correlation coefficient

Spreading dynamics is the most common process in many domains, such as physics and society. We utilize the SIR model to simulate the spreading process for evaluating the effectiveness of the WH on quantifying spreading influence. We apply the Kendall's tau $\tau$ correlation coefficient to evaluate the prediction accuracy. The greater absolute value of $\tau$ implies higher correlation between two sample vector. That means higher correlation between the WH value vector and the spread range vector indicates better prediction accuracy. The Kendall's tau $\tau$ between the node influence index $\sigma$ of SIR model and six centralities indices is summarized in Table 3. One can observe that the proposed method which is highly correlated with the size of the infected population of the SIR model outperforms the other ranking means in most networks. It is worth mentioning that the results are very similar which can be found in the Table 4 ($\beta = 1.5\beta_c$) and Table 5 ($\beta = 2\beta_c$). The rank correlation of WH may not has the highest value among all centralities in a small value of $\beta$ like Oregon. But the proposed method acquires the best performance in higher spreading probability $\beta$ in Table 4 and Table 5.

Table 3 The Kendall Tau $\tau$ correlation coefficient between the node influence vector simulated by SIR model and different centrality index vector. The spreading rate $\beta$ can be found in table 2. In each row, the largest $\tau$ is highlight in bold.

| Network | $\tau(D,\sigma)$ | $\tau(B,\sigma)$ | $\tau(C,\sigma)$ | $\tau(kS,\sigma)$ | $\tau(H,\sigma)$ | $\tau(WH,\sigma)$ |
|---|---|---|---|---|---|---|
| Karate club | 0.68650 | 0.56047 | 0.66500 | 0.62465 | 0.67650 | **0.82068** |
| Dolphins | 0.79718 | 0.53940 | 0.57881 | 0.71288 | 0.82011 | **0.84768** |
| Jazz | 0.74026 | 0.45134 | 0.65563 | 0.7440 | 0.78381 | **0.80102** |
| USAir | 0.69820 | 0.54173 | 0.72679 | 0.71651 | 0.71231 | **0.77069** |
| C. elegans | 0.57288 | 0.43609 | 0.53407 | 0.59691 | 0.58204 | **0.62888** |
| Email | 0.72219 | 0.58623 | 0.72356 | 0.74864 | 0.74826 | **0.78680** |
| Facebook | 0.68180 | 0.44910 | 0.34362 | 0.71346 | 0.70742 | **0.75747** |

| Network | | | | | | |
|---|---|---|---|---|---|---|
| Power grid | 0.60196 | 0.42367 | 0.30214 | 0.51417 | 0.61773 | **0.80600** |
| GrQc | 0.70858 | 0.42658 | 0.56805 | 0.68711 | 0.71714 | **0.75446** |
| HepTh | 0.58417 | 0.45521 | 0.71238 | 0.60185 | 0.61067 | **0.76169** |
| PGP | 0.60274 | 0.41604 | 0.49540 | 0.57068 | 0.60511 | **0.65661** |
| Oregon | **0.37469** | 0.25616 | 0.26758 | 0.37163 | 0.37333 | 0.28735 |
| HepPh | **0.71833** | 0.40116 | 0.56480 | 0.69728 | 0.70924 | 0.63276 |
| AstroPh | 0.70169 | 0.43844 | 0.71343 | 0.70445 | 0.71552 | **0.76860** |
| CondMat | 0.61579 | 0.38415 | 0.67564 | 0.63366 | 0.64324 | **0.75636** |
| Enron | 0.52997 | 0.45444 | 0.45154 | 0.53003 | **0.53342** | 0.47791 |

Table 4 The Kendall Tau $\tau$ correlation coefficient between the node influence vector simulated by SIR model and different centrality index vector.. The spreading rate $\beta$ is set as $\beta = 1.5\beta_c$. In each row, the largest $\tau$ is highlight in bold.

| Network | $\tau(D,\sigma)$ | $\tau(B,\sigma)$ | $\tau(C,\sigma)$ | $\tau(kS,\sigma)$ | $\tau(H,\sigma)$ | $\tau(WH,\sigma)$ |
|---|---|---|---|---|---|---|
| Karate club | 0.74690 | 0.60650 | 0.67971 | 0.65700 | 0.72534 | **0.79138** |
| Dolphins | 0.69303 | 0.46891 | 0.64165 | 0.69937 | 0.76463 | **0.77986** |
| Jazz | 0.78788 | 0.45985 | 0.69161 | 0.78595 | 0.83429 | **0.86103** |
| USAir | 0.71178 | 0.55152 | 0.74683 | 0.73731 | 0.73279 | **0.81158** |
| C. elegans | 0.57587 | 0.41369 | 0.56223 | 0.61404 | 0.58669 | **0.68422** |
| E-mail | 0.77379 | 0.61714 | 0.77381 | 0.79641 | 0.80500 | **0.86005** |
| Facebook | 0.62200 | 0.4251 | 0.40978 | 0.66601 | 0.65256 | **0.78746** |
| Power grid | 0.42410 | 0.29214 | 0.39183 | 0.39870 | 0.46459 | **0.68928** |
| GrQc | 0.63079 | 0.39253 | 0.64386 | 0.62729 | 0.64840 | **0.78230** |
| HepTh | 0.56251 | 0.45726 | 0.80169 | 0.59146 | 0.59762 | **0.81910** |
| PGP | 0.51525 | 0.35000 | 0.58276 | 0.51175 | 0.52874 | **0.70986** |
| Oregon | 0.38280 | 0.23639 | 0.34603 | **0.38830** | 0.38472 | 0.35581 |

Table 5 The Kendall Tau $\tau$ correlation coefficient between the node influence vector simulated by SIR model and different centrality index vector. The spreading rate $\beta$ is set as $\beta = 2\beta_c$. In each row, the largest $\tau$ is highlight in bold.

| Network | $\tau(D,\sigma)$ | $\tau(B,\sigma)$ | $\tau(C,\sigma)$ | $\tau(kS,\sigma)$ | $\tau(H,\sigma)$ | $\tau(WH,\sigma)$ |
|---|---|---|---|---|---|---|
| Karate club | 0.63741 | 0.54327 | 0.66263 | 0.62667 | 0.67004 | **0.72027** |
| Dolphins | 0.71113 | 0.49388 | 0.66863 | 0.72416 | 0.76153 | **0.78125** |
| Jazz | 0.81011 | 0.46428 | 0.68955 | 0.79323 | 0.84709 | **0.86789** |
| USAir | 0.72417 | 0.54954 | 0.77476 | 0.75250 | 0.74583 | **0.86298** |
| C. elegans | 0.62053 | 0.45687 | 0.54898 | 0.66351 | 0.63958 | **0.72436** |
| E-mail | 0.81928 | 0.64867 | 0.77756 | 0.83057 | 0.74755 | **0.87126** |
| Facebook | 0.63007 | 0.42532 | 0.44406 | 0.67938 | 0.66365 | **0.81913** |
| Power grid | 0.31547 | 0.20293 | 0.46989 | 0.30782 | 0.35624 | **0.52467** |
| GrQc | 0.57436 | 0.37810 | 0.71264 | 0.57912 | 0.59853 | **0.79230** |
| HepTh | 0.59150 | 0.47677 | 0.80892 | 0.62076 | 0.63027 | **0.84470** |
| PGP | 0.47282 | 0.31974 | 0.63765 | 0.48665 | 0.49155 | **0.72975** |
| Oregon | 0.38756 | 0.20486 | 0.40210 | 0.39803 | 0.39154 | **0.40692** |

Next, we investigate the Kendall's tau $\tau$ correlation coefficient between the node influence index $\sigma$ used SIR model and six centrality indices by varying the infection probability $\beta$ from 0.01 to 0.2 in eight actual networks. The calculations, to evaluate the effect of infection probability $\beta$, is shown in Fig.4. The WH exhibits obviously correctness on a wide range of probabilities $\beta$ in eight real networks, especially when the infection probabilities $\beta$ is around the epidemic threshold $\beta_c$ (the black dot line). When the infection probabilities $\beta$ is very small, the spreading is typically confined to the neighborhood of the initially infected node, hence the node with larger degree can infected more nodes. That is why D always achieve the largest $\tau$ values when $\beta$ is less than the epidemic threshold $\beta_c$ for Karate Club, Dolphins, GrQc and Email. When $\beta$ become larger, our method begins to show better performance. Although WH becomes less effective than D, KS and H centralities when $\beta$ grows much larger than $\beta_c$ in Jazz, US Air, C.elegans and Email. The proposed method can still achieve a better performance on a wide of $\beta$. So the above mentioned results demonstrated that the WH has a better indicator to identify the spreading influence in complex networks whenever the infection probability $\beta$ is greater than the epidemic criticality $\beta_c$.

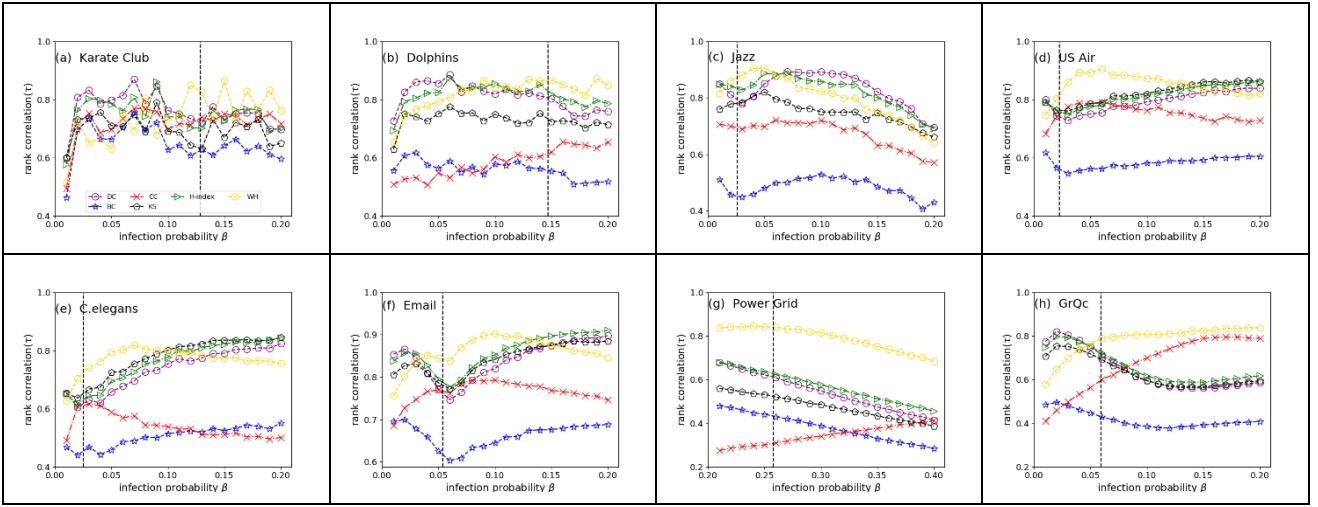

Fig.4. the Kendall's tau $\tau$ correlation coefficient between the node influence index simulated by SIR model and six centrality indices is plot by varying the infection probability $\beta$ in eight actual networks. The real-world networks are (a) Karate Club, (b) Dolphins, (c) Jazz, (d) US Air, (e) C.elegans, (f) Email, (g) Power grid and (h) GrQc.

### 3.3 Monotonicity and efficiency

The monotonicity $M$ is defined to quantify the fraction of ties in the ranking list. The higher the value of $M$ means the ranking method has a better resolution of different influential nodes. The monotonicity $M$ of different ranking methods is summarized in Tabled 6. The WH and C are the best measure of the six measures. In order to clarify the ranking distribution, we plot a complementary cumulative distribution function (CCDF) as shown in Fig. 5.

Table 6 the monotonicity $M$ of different methods was applied to sixteen real world networks. The value of $M(\cdot)$ is the monotonicity of the corresponding measures.

| Network | $M(D)$ | $M(B)$ | $M(C)$ | $M(KS)$ | $M(H)$ | $M(WH)$ |
| --- | --- | --- | --- | --- | --- | --- |
| Karate club | 0.7079 | 0.7754 | 0.8993 | 0.4958 | 0.6012 | **0.9507** |
| Dolphins | 0.8312 | 0.9623 | 0.9737 | 0.3769 | 0.7017 | **0.9937** |
| Jazz | 0.9659 | 0.9886 | 0.9878 | 0.7944 | 0.9398 | **0.9993** |
| US Air | 0.8586 | 0.6970 | 0.9892 | 0.8114 | 0.8467 | **0.9945** |
| C. elegans | 0.7922 | 0.8743 | 0.9900 | 0.6962 | 0.7599 | **0.9961** |
| Email | 0.8874 | 0.9400 | 0.9988 | 0.8088 | 0.8661 | **0.9996** |
| Facebook | 0.9740 | 0.9855 | 0.9967 | 0.9492 | 0.9674 | **0.9998** |
| Power Grid | 0.5927 | 0.8322 | **0.9998** | 0.2560 | 0.4776 | 0.9606 |
| GrQc | 0.7460 | 0.3823 | **0.9858** | 0.6631 | 0.7126 | 0.9835 |
| HepTh | 0.7627 | 0.5077 | **0.9931** | 0.6742 | 0.7297 | 0.9908 |
| PGP | 0.6193 | 0.5123 | **0.9996** | 0.4806 | 0.5836 | 0.9920 |
| Oregon | 0.4940 | 0.4301 | **0.9936** | 0.4399 | 0.4801 | 0.9935 |
| HepPh | 0.8657 | 0.5246 | **0.9978** | 0.8248 | 0.8483 | 0.9976 |
| AstroPh | 0.9264 | 0.6361 | **0.9990** | 0.9082 | 0.9175 | **0.9990** |
| CondMat | 0.8524 | 0.4922 | **0.9980** | 0.7980 | 0.8268 | 0.9974 |
| Enron | 0.7370 | 0.3392 | 0.9936 | 0.7141 | 0.7259 | **0.9989** |

The D, KS and H are decrease rapidly because many nodes are in the same ranking value. And the CCDF of WH and C decline down in different types of networks. This illustrates that by combining the previous analyzing, the WH is more efficient to identify the influential nodes, although the WH and C have an excellent performance at distinguishing the spreading capability of the influential nodes. Interestingly, the CCDF of B centrality has a phenomenon of dropping down. Although the CCDF plot of B centrality decline down more slowly than the others in many networks, B has a worse performance of monotonicity than WH and C.

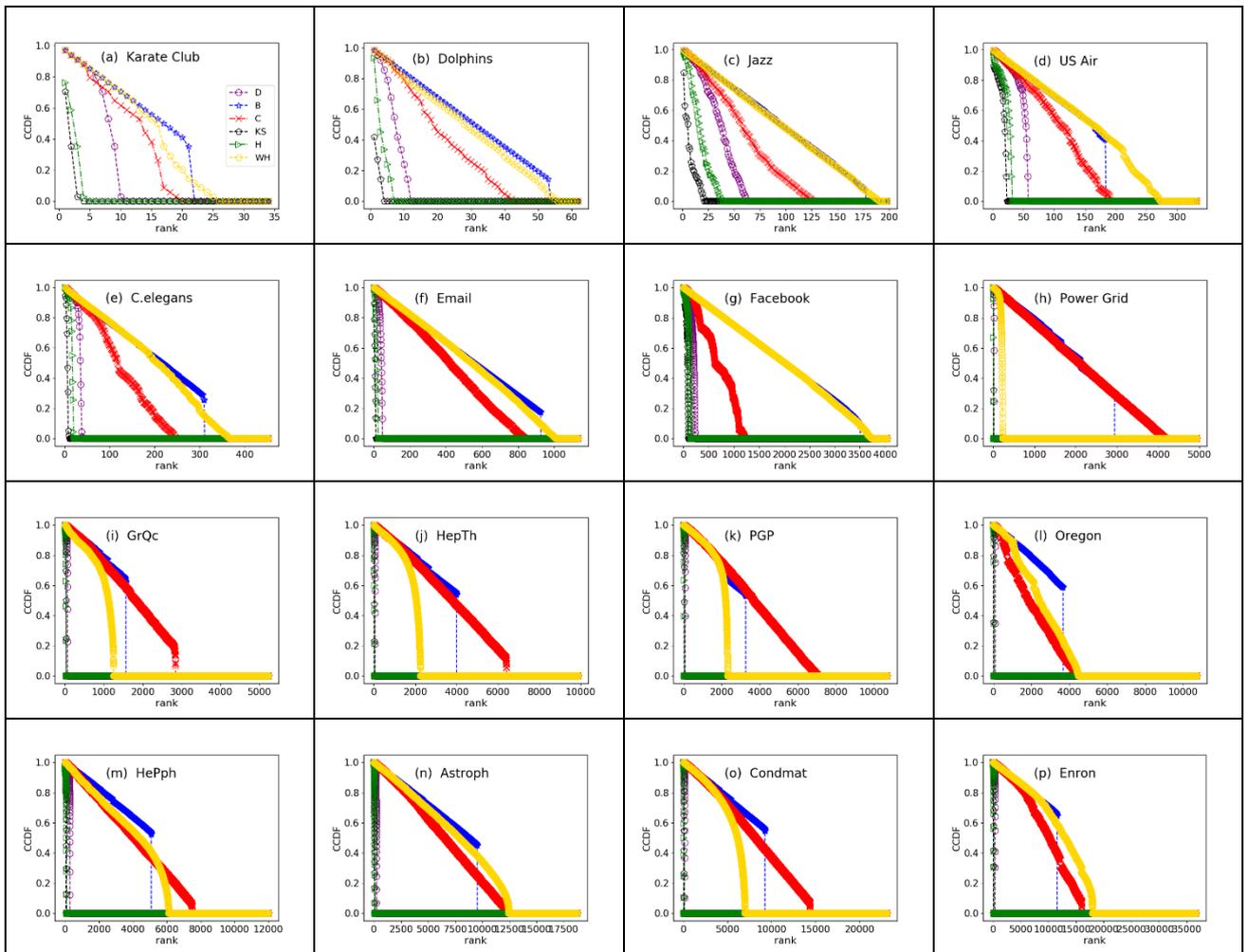

Fig.5. The distribution of ranks in sixteen real networks. The real-world networks are (a) Karate club, (b) Dolphins, (c) Jazz, (d) USAir , (e) C.elegans, (f) E-mail, (g) Facebook, (h) Power Grid, (i) GrQc, (j) HepTh, (k) PGP, (l) Oregon, (m) HepPh, (n) AstroPh, (o) CondMat and (p) Enron. The CCDF plot of D, KS and H drops quickly, whereas the distribution of WH and C decrease monotonically in most case. The CCDF plot of B has a phenomenon of dropping down.

In this part, the comparison of the computing complexity for the proposed method and other five measure is discussed. The six methods' computational complexity is shown in Table. 7, where n is the total number of nodes, m is the number of edges in a network.

The proposed method of calculating the WH has a complexity of $O(m)$, given the degree of each node is known. In addition Different methods have different computation complexity and require different network information.

Both the Floyd's algorithm [3_1] and Brandes' algorithm [3_2] are used to count the number of shortest paths. Calculating the shortest paths can be done using Floyd's algorithm search in time $O(n^3)$ and Brandes' algorithm [3_2] in time $O(mn)$. The computing complexity of C can be computed in time $O(n^3)$ using Floyd's algorithm [3_1]. The closeness centrality takes $O(n^2 \log n + nm)$ when calculated in a sparse graph using Johnson's algorithm [3_3]. The D, B and H only need the local information of a node and the KS needs global information. The computational complexity of D, B, H and KS is $O(m)$, which indicates that the proposed method has a lower computation efficiency and can be used in large-scale network.

Table 7 the computational complexity of six methods.

| Method | Information | Computational complexity |
|---|---|---|
| D | Local information | $O(m)$ |
| B | Global information | $O(nm)$ or $O(n^3)$ |
| C | Global information | $O(n^2 \log n + nm)$ or $O(n^3)$ |
| KS | Global information | $O(m)$ |
| H | Local information | $O(m)$ |
| WH | local information | $O(m)$ |

Therefore our measure provides an effective way to rank the influential nodes without no increase in complexity than other time consuming measure. So we can argue that the WH is efficient and accurate for identifying and ranking the influential nodes.

## 4. Discussion

Spreading like epidemic and information is a ubiquitous process in the social, biological and technological networks. Therefore, identifying influential nodes, which can optimize and conserve spreading resources in a large scale of complex networks, is full of theoretical and practical significance.

In this paper, we proposed a novel method: the weighted h-index centrality to identify and rank the spreading ability of nodes in complex networks. This measure collects centrality information by adding together the weighted h-index of neighbors. To evaluate the performance, we apply the proposed method on sixteen actual networks compared with the size of the infected population in the SIR model. We find that the WH outperforms the other five methods by employing the impression function $(\varepsilon)$ and the Kendall's tau $(\tau)$ correlation coefficient to measure the rank imprecision and correlation. The proposed weighted h-index centrality exhibits obviously effectiveness of lower imprecision, higher $\tau$ and more competitive monotonicity compared to the other methods. Moreover, we analyze the computational complexity of six methods and the results show that our method is more suitable for large-scale networks because of using local information and having a lower complexity. Therefore, the proposed method can offer an excellent performance on discriminating the influence capability of nodes and provide a more reasonable and efficient ranking in complex networks.

Here, the proposed method has a good extensibility, as our algorithm is based on a node's nearest neighbors. We only concentrate on the centrality information of a node undirected networks, some extensions applying our method may be worth studying as in Ref [40, 41, and 42,1_4,1_12,1_13,1_14,1_15]. Recently, some researches began to concern diverse structure of networks and spreading dynamics, which play a significant role in the communication behavior [43, 44,4_1]. So further work is to apply and find more efficient method to identify and rank the node spreading influence using our method with the mentioned issues.

## Acknowledgments

The authors thank for support from the National Natural Science Foundation of China (No.71571017, No.91646124 and No.71621001), and support from the Fundamental Research Funds for the Central Universities (2017YJS102, 2015JBM058).